\documentclass[preprint]{aastex}

\shorttitle{AGNs in the Local Universe}
\shortauthors{Maia, Machado \& Willmer}
\begin{document}

\title{The Seyfert Population in the Local Universe\altaffilmark{1}} 
\author{Marcio A. G. Maia and Rodolfo S. Machado}
\affil{GEA - Observat\'orio do Valongo - UFRJ
and Observat\'orio Nacional/MCT, Rio de Janeiro~-~RJ, Brazil}
\email{maia@ov.ufrj.br, rodolfo@on.br}

\author{and}

\author{Christopher N. A. Willmer\altaffilmark{2}}
\affil{UCO/Lick Observatory, University of California, Santa Cruz, 95064}

\email{cnaw@ucolick.org}

\altaffiltext{1}{Partly based on observations at European Southern Observatory 
(ESO), under the ESO-ON agreement to operate the 1.52m telescope.}

\altaffiltext{2}{on leave from Observat\'orio Nacional/MCT, Rio de Janeiro~-~RJ, Brazil}


\begin{abstract}

The magnitude-limited catalog of the Southern Sky Redshift Survey (SSRS2), is used to 
characterize the properties of galaxies hosting Active Galactic Nuclei. Using 
emission-line ratios, we identify a total of 162 (3\%) Seyfert galaxies out of the 
parent sample with 5399 galaxies. The sample contains 121 Seyfert 2 galaxies and 41 
Seyfert 1. The SSRS2 Seyfert galaxies are predominantly in spirals of types Sb and 
earlier, or in galaxies with perturbed appearance as the result of strong interactions 
or mergers. Seyfert galaxies in this sample are twice as common in barred hosts than 
the non-Seyferts.  By assigning galaxies to groups using a percolation algorithm we 
find that the Seyfert galaxies in the SSRS2 are more likely to be found in binary 
systems, when compared to galaxies in the SSRS2 parent sample.  However, there is no
statistically significant difference between the Seyfert and SSRS2 parent sample when
systems with more than 2 galaxies are considered.
The analysis of the present sample suggests that there is a stronger 
correlation between the presence of the AGN phenomenon with internal properties of 
galaxies (morphology, presence of bar, luminosity) than with environmental effects 
(local galaxy density, group velocity dispersion, nearest neighbor distance).

\end{abstract}



\section{Introduction}

Since bright AGNs are such a rare phenomenon, the way a sample is
selected can affect the interpretation of the observational results.  
As an example, the largely accepted unified model for AGNs is sometimes challenged 
by the results of some analyses that may be dependent on the sample selection 
\citep [e.g.,][]{Lau95}.  Seyferts selected on the basis of X-ray or UV 
excess are biased in favor of the type 1.  
On the other hand, samples selected in the IR are prone to contain galaxies 
undergoing strong star formation activity, and also biased in favor of luminous 
Seyfert nuclei.  The FIR correlates weakly with the AGN emission and is very likely 
related to a concurrent starburst.
The identification of large numbers of galaxies hosting Active Galactic Nuclei has 
been possible thanks to the use of spectroscopic surveys using objective prisms 
(e.g., \cite{Sal00},KISS; \cite{Gro01} and references therein). However, since 
these surveys are specifically designed to detect galaxies with emission-lines, using 
their catalogues to compare the properties of AGN host galaxies with the general 
field galaxies is not straightforward. The need of an isotropic property to assure 
a fair selection is therefore very important. 

The use of magnitude-limited surveys is an effective means of selecting galaxies in 
a relatively unbiased way, such that detailed comparisons between the properties of 
AGNs and the entire sample can be done. The first analysis of this kind was by 
\citet{Huc92}, who used the CfA1 spectroscopic database to identify AGN hosts, and 
calculated their luminosity function to determine their density relative to the entire 
sample of galaxies.  \citet{Huc92} found that AGNs dominate the bright end of the CfA1
survey luminosity function.  A similar conclusion was reached by \citet{Koh97} who 
used the Hamburg-ESO prism survey, and found that the spatial densities of Seyfert 1 
galaxies at $z<$0.07 are consistent with those of \citet{Huc92}.
A detailed analysis using a smaller sample than the CfA1 was done by \citet{Ho97b} 
and \citet{Ulv01} who selected galaxies brighter than $B_T$ = 12.5, for which high 
resolution (and high signal-to-noise) spectroscopy was obtained.  \citet{Ho97b} and 
\citet{Ulv01} showed that $\sim$ 40\% of all galaxies present AGN-like spectra, about 
2/3 of the sample being LINERS, while a proportion of  50\% to 75\%  of all AGN
are found in bulge-dominated galaxies. 

The largest sample of AGNs to date is those compiled by \citet{Hao03}
and \citet{Kau03}, both works using the Sloan Digital Sky Survey to detect AGNs. 
Using the H$\alpha$ flux as an estimator of the energy output from the AGN, 
\citet{Hao03} find that the ratio between Seyfert 1 and Seyfert 2 galaxies is 
roughly the same at low luminosities, while at higher luminosities, Seyfert 1 
galaxies dominate by a large margin.  In that work they find that there is no 
correlation between the nuclear luminosity and the host galaxy luminosity. 

In the present work we examine the properties of AGN hosts relative to the population 
of field galaxies. For this we use a sample of AGNs identified in the SSRS2 survey 
\citep{daC98}.  This survey reaches 
about 1 magnitude deeper than the CfA1 and contains more than twice the number
of galaxies.  In this work, the SSRS2 is used as a ``control sample'' against 
which several properties of the AGN hosts are compared, therefore allowing a 
systematic characterization of the properties of galaxies relative to the general 
field galaxy sample. This paper is divided as follows: in Section 2 we describe the 
sample of galaxies, followed in Section 3 several analyses of the properties of AGN 
hosts as a function of the internal properties of galaxies. In Section 4 we examine 
how the AGN phenomenon correlates with the environment where the host galaxy is 
located. A summary of our results is presented in Section 5. 

In the present paper we use the following cosmological parameters:
$H_0$ = 70 kms$^{-1}$ Mpc$^{-1}$,  $\Omega_m$ = 0.3, $\Lambda$ = 0.7


\section{The Sample}

To characterize the AGN properties in the nearby Universe, we used an updated 
version of the Southern Sky Redshift Survey \citep[SSRS2,][]{daC98}.  
This catalog contains 5399 galaxies with $m_B \le 15.5$ covering 1.69 sr of the 
southern sky, limited in the southern galactic hemisphere by  
$-40^{o}\le \delta \le -2.5^{o}$ and $b \le -40^{o}$, and in the northern galactic 
hemisphere by $\delta \le 0^{o}$ and $b \ge +35^{o}$.  
This updated version of SSRS2 contains more precise positional information, new or 
improved radial velocities, revised morphological classifications, and has had removed 
galaxies with magnitudes above the nominal limit as well as false detections in the 
HST Guide Star Catalogue. It also contains galaxies that where not included previously 
because of misclassification. This catalog is currently 99.99\% complete in redshift,  
and in our database we have optical spectra available for $\sim$ 68\% of the SSRS2 
galaxies, while the remaining radial velocities were obtained from the literature.  

The spectra were inspected and AGNs identified using the diagnostic diagrams of line 
intensity ratios proposed by \citet{BPT} and \citet{Vei87}. We classify as Seyfert 
galaxies objects that satisfy: 
$$ {[NII]\lambda 6584 \enskip / \enskip H\alpha} \enskip > \enskip 0.6 $$
\noindent and 
$$ {[OIII]\lambda 5007 \enskip / \enskip H\beta} \enskip > \enskip 3 $$  
\noindent In dubious cases, we also used the following ratios whenever available:
$$ {[OII]\lambda 3727 \enskip / \enskip [OIII]\lambda 5007} \enskip < \enskip 1 $$
\noindent and/or
$$ {[OI]\lambda 6300 \enskip / \enskip H\alpha } \enskip < \enskip 0.3 $$   

We classify as Seyfert 1 those galaxies with broad $H\alpha$ lines in their 
spectra.   Using the prescriptions described above, we found 98 Seyfert galaxies.  
Some of them were detected in the SSRS2 survey and were reported in the papers by 
\citet{Mai87} and \citet{Mai96}.  In cases where the Seyfert nature was dubious due 
to low signal to noise or small wavelength coverage of the spectra, new spectroscopic 
data were obtained. 

Additional Seyfert galaxies in the region covered by the SSRS2 were included after 
searching the NED database and the literature. The literature sometimes presents 
conflicting classifications, or does not state clearly how galaxies were classified. 
A very common confusion is to classify LINERS as Seyferts.  For all candidate
objects recovered from the literature, we obtained spectra,  and measured the line 
ratios so their Seyfert nature could be confirmed.  This ensures that the present 
catalog has a homogeneous classification procedure.  A total of 64 additional galaxies 
were added to our list. 

The catalog of SSRS2 Seyfert galaxies (hereafter, AGNC) is presented in Table 1, which 
contains the object name, equatorial coordinates, apparent magnitude, morphological 
type, axial ratio $b/a$, the heliocentric radial velocity and the Seyfert type (1 or 2).

The apparent magnitudes in Table 1 were derived from the HST Guide Star Catalogue 
\citep{Las90}, and refer to the galaxy as a whole, there being no decomposition between 
the contribution from the host galaxy and the AGN.  The tabulated magnitudes are shown 
by \citet{Alo94} to be isophotal magnitudes with a limiting isophote close to 26 mag 
arc~sec$^{-2}$; these magnitudes are also very close to the $B(0)$-Zwicky magnitude 
system used in the CfA1 survey. The conversion between $B(0)$-Zwicky and total 
magnitudes is given by $B(0)$-Zwicky = $B_T$ + 0.26  \citep{Fel85}. 

In the subsequent analyses, the observed magnitudes are corrected for galactic 
extinction using the DIRBE/IRAS maps \citep{Sch98}, and using $A/E(B-V)$ = 4.035, 
where we assume that the bandpass of the plates used for the GSC is the same as that
used for the APM survey \citep{Mad90}. In the analyses, the radial velocities of 
galaxies are corrected to the Local Group Barycenter following \citet{Yah77}.  

The absolute magnitudes are calculated using the expressions presented by 
\citet{Hog00}. For the K-corrections we use morphological-type dependent 
expressions due to \citet{Pen76}.

In Figure 1 we show the distribution of apparent magnitudes of galaxies in the AGNC 
and the SSRS2. This figure suggests there might be a lack of AGNs for magnitude bins 
fainter than m$_B$=15.0. This possible incompleteness is also suggested by the 
$V/V_{max}$ statistic \citep{Sch68}, which for the SSRS2 sample is 0.47, while for 
the AGNs, it is 0.38. 

An independent assessment of the catalogue completeness can be made using the 
completeness test ($T_c$) proposed by \citet{Rau01}, shown in Figure 2. In this test, 
the completeness is estimated by calculating for each galaxy $i$ the ratio between 
the number of galaxies with absolute magnitudes $M \leq M_i$ ($r_i$) and the number 
of galaxies ($n_i$) visible to survey limit ($m_{lim}$) in both cases within the
volume out to $z = z_i$. From these one calculates the estimator 
$$ \bar \zeta = \frac {r_i} {n_i+1} $$
which is related to the ratio between the cumulative luminosity function
$$ \zeta = \frac {F(M\leq M_i)} {F(M \leq M_{lim}(z))}  $$
The expectation value and variance of $\zeta$ are respectively :
$$  E_i = \frac {1} {2} $$
and
$$  V_i = \frac {1} {12} \frac {n_i-1} {n_i+1}.  $$

$T_c$ is then defined as the ratio between the fluctuations of the
cumulative luminosity function and its variance:
$$T_c = \frac { \sum_{i=1}^{N_{gal}} (\bar \zeta -  \frac {1} {2})} { \biggl\lbrace
\sum_{i=1}^{N_{gal}} V_i \biggl\rbrace^{\frac {1} {2}}}  $$
(see \citet{Rau01} for the full derivation). The underlying assumption that is made 
in this test is  that there is no dependence between an object's luminosity and 
its location. 
This test shows that the sample completeness cannot be rejected at a confidence level 
greater than 95\%, suggesting that the incompleteness of the present sample may not 
be severe at the adopted limiting magnitude ($m_B$ =15.5).

In Figure 3 we show the radial velocity distribution of AGNC and SSRS2 galaxies. For 
radial velocities $\ge$ 13,000 km s$^{-1}$ we see that both samples present few 
galaxies, this velocity range being populated by galaxies about 1.5 magnitudes 
brighter than $M^*$. The Kolmogorov-Smirnov test (KS), applied to the SSRS2 and AGNC
samples shows that their velocity distributions differ at the 80\% confidence level. 
This may be reflecting the fact that bright AGNs will contribute significantly to the 
total emission of galaxies, therefore enhancing the Malmquist bias, such that brighter 
and more distant galaxies will be included in the sample.
The average distance, $<d>$, of the galaxies in the samples are $<d_{SSRS2}>=108$ Mpc 
and $<d_{AGNC}>=100$ Mpc.

We show in Figure 4 the fraction of available spectra as a function of galaxy 
morphology for galaxies in the SSRS2, where the galaxy morphologies are coded 
following \citet{deV76}.  The SSRS2, in general, samples spectroscopically all 
morphologies at a similar rate, with exception of T=-1 (S0a) and very late types, 
such as Sds and Irregular galaxies. This figure can also be used to estimate the 
fraction of AGNs that may have been missed in the SSRS2 because no spectra are 
available. 
By considering the fraction of galaxies with spectra in the SSRS2 (68\%) database,
and the fraction of AGNs detected in each morphological type, the expected number of
active galaxies in the 32\% without previous spectra in our database should be of 
about 42. This is less than the 64 galaxies recovered from the inspection of the 
literature and may be an indication that the AGNC is not drastically affected by 
the lack of spectra for 32\% of SSRS2 galaxies.


\section{Properties of AGN host galaxies}

In this section we use the AGNC to characterize the properties of Seyfert galaxies, 
and compare them to a representative ensemble of galaxies of the Local Universe

The total number of Seyfert galaxies identified in the SSRS2 is 162. This comprises 
about 3\% of the total number of galaxies in the sample (5399).  This number is 
consistent with the proportion of 2\% found by \citet{Huc92} for the 48 Seyferts of 
the CfA1, the 4.7\% found by \citet{Hao03} for the SDSS and the $\sim$ 4\% found for 
KISS by \citet{Gro01}.
We should note  that \citet{Mai95} find that the Revised Shapley-Ames Catalog of Bright 
Galaxies \citep[RSA][]{San81}, contains 5\% (91 galaxies) of AGNs, although they claim, 
using completeness arguments, that AGNs may comprise up to 16\% of the sample. This is 
supported also by \citet{Ho97a} who find that about 11\% of the galaxies in a
subsample of 486 nearby galaxies selected from the RSA catalog are Seyferts.
In a mid-IR selected sample of 891 galaxies, \citet{Hun99} find a fraction of 9\% 
of AGNs (34 type 1 and 44 type 2).

Part of these differences may be explained by the criteria used in different works 
to classify a galaxy as an AGN. In particular, IRAS samples are strongly biased 
towards luminous, dusty, early-type spiral galaxies.  We should note that the 
present sample of AGNs identified from the SSRS2 may be deficient in low-luminosity 
AGNs, particularly when compared to works like \citet{Ho97a} who used a different 
strategy that allowed uncovering lower luminosity AGNs, through the use of 
high-dispersion spectroscopy \citep{Ho97a}.

In the AGNC catalogue, which contains 162 galaxies, there are 121 Seyfert type 2 
galaxies and 41 type 1, corresponding to a ratio of $\sim$ 3:1. For the KISS, the 
proportion is $\sim$ 2:1 \citep{Gro01}, while the ratio between narrow-line 
and broad-line AGNs in the SDSS is also 2:1 \citep{Hao03}.
Other authors that measured this proportion found ratios of 4:1 for the RSA 
\citep{Mai95}, 1:1 for the CfA1 \citep{Huc92}, and 1:1 found by \citet{Rus93} for 
their 12 $\mu$m flux-limited sample of galaxies.  

\subsection{Morphology}

We display the fraction of galaxies as a function of morphological type in Figure 5. 
The AGNC (solid line) and SSRS2 (dashed line) distributions differ at a confidence 
level of 99.99\%. The Seyferts are distributed preferentially among early-type spirals, 
confirming previous results \citep[e.g.,][]{Mol95, Hun99, Gro01}, with 70\% of the AGNs 
between the morphological types S0a and Sb, there being $\sim$ 5\% of AGNs in 
ellipticals and very few hosted by spirals later than Sb. Similar results are found 
by \citet{Kau03} in their analysis of SDSS data. Although SDSS has no morphologies, 
\citet{Kau03} were able to show that most of the AGN hosts in their data are found 
in centrally concentrated galaxies, similar to early type galaxies.
There is a high proportion of Seyferts in galaxies with type 15, which contain about 
10\% of the total sample of AGNs. This morphological type is used in the SSRS2 to 
denote galaxies presenting evidence of advanced merger event.

Evidence of segregation in terms of morphologies was presented by \citet{Hun99}, who 
claimed that Seyfert 1s are more often found in earlier spirals compared 
to Seyfert 2s. By analysing  the KISS galaxies \citet{Gro01} find that Seyfert 2 
(and LINERS) are preferentially found in redder hosts ($B-V$ = 0.92), while Seyfert 1 
hosts typically have bluer colors ($B-V$=0.70). 
This apparent contradiction between the two works can be explained by the use of 
integrated colors of galaxies i.e., the contribution of the AGN is
very hard to be measured separately, added to the fact that the correlation between 
galaxy morphology and color has a large dispersion. Yet both works suggest that 
the Seyfert types are preferentially found in different types of host.
The AGNC was used to investigate whether this segregation in morphologies between the 
Seyfert 1 and 2 hosts is found in the SSRS2 (Figure 6). By using a KS test we find 
that both samples may be considered  identical at the 84 \% confidence level,
showing that for the SSRS2 there is no strong correlation between the host
galaxy morphology and the Seyfert type.


\subsection{Luminosities}

The distribution of absolute magnitudes, $M_B$ for the AGNC and SSRS2 catalogs is 
displayed in Figure 7 panel (a), and between the Seyfert types 1 and 2 in panel (b). 
Panel (a) clearly shows that galaxies hosting AGNs have a peaked distribution, while 
the non-actives present a tail at lower luminosities. 
The mean values of $M_B$ are: -20.06 $\pm$ 1.32 for the SSRS2, -20.59 $\pm$ 0.87 for 
AGNs, -20.85 $\pm$ 0.77 for type 1 and -20.50 $\pm$ 0.89 for type 2. 
The evidence here is that AGNs reside preferentially in high luminosity hosts, as
already suggested by \citet{Huc73} and confirmed by \citet{Huc92}, and
 \citet{Gro01}. Since we cannot isolate the light from the AGN from that 
of the host with the present data,  it is unclear how much the observed distribution 
is due to the host galaxy and how much is due to the AGN itself. As shown by 
\citet{Lon00}, the correlation between the $total$ and typical $nuclear$ 
luminosities of galaxies with AGNs as measured from an HST-selected sample, is fairly 
tight for high-luminosity AGNs (rms $\sim$1 mag), but shows a broad dispersion 
once lower-luminosity AGNs are considered (rms $>$1 mag). 
This may be explained by the difficulty in isolating the contribution from the AGN 
even in the case of HST data.  This interpretation is supported  by \citet{Hao03} 
who show using SDSS data, that there is no correlation between the AGN luminosity 
and that of the host galaxy. Notwithstanding, the tendency we have found may be 
attributed, at least in part, to the relation between the black hole
mass  and that of the galaxy bulge.


\subsection{Axial Ratios}

Since Seyfert hosts are frequently found in spiral galaxies, it is possible that 
optically-selected samples may be biased against identifying AGNS in edge-on systems 
by projection effects. This is because the gas which is ionized by the nucleus is not 
detected along the line of sight, being obscured by the intervening gas and dust in 
the plane of the galaxy.  The distribution of the axial ratio $b/a$ for the
non-interacting spirals of the AGNC is similar to that of a magnitude-limited sample 
of thin circular disks \citep{Mai95}. The frequency distribution of axial ratios 
for all Seyferts is presented in Figure 8 panel (a).
The lack of objects with very small axial ratios $b/a$ is expected because of the disk 
thickness and also to the less numerous edge-on systems in magnitude-limited samples. 
The small decrease in frequencies for $b/a$ above 0.8 may be attributed to the 
existence of noncircular disks.  

In panel (b) of Figure 8 we show the distribution of axial ratios for Seyfert 2
galaxies (solid line) and Seyfert 1 (dashed line). Seyfert 2s tend to be more abundant 
at smaller values of $b/a$ while Seyfert 1s are commoner at $b/a$ closer to 1. 
A KS test shows that both samples differ at the 97\% confidence level.  This behavior 
is consistent with the interpretation that part of Seyfert 2 galaxies are Seyfert 1s 
heavily obscured in more inclined systems.  

\subsection{Presence of Bar}

Bars are claimed to be detected in up to 50\% of normal galaxies 
\citep[e.g.,][]{Sel93,Kna00}, although some authors claim that the fraction may be as 
high as $\approx$70\%  \citep[e.g.,][]{Hun99}. The presence of this non-axisymmetric 
gravitational perturbation can induce radial gas inflow, fueling the Seyfert  
activity as described by \citet{Hel94}. 
The incidence of bars as a function of morphological type was reported by 
\citet{Ho97b}, although \citet{Kna00} do not find any trend.  \citet{Lai01} using 
IR HST data, find that bars are more abundant in Seyfert hosts (73\%) than in 
non-Seyferts (50\%). On the other hand, \citet{Mol95}, \citet{Mul97} and \citet{Mal98} 
find similar occurrence of barred systems in Seyferts and normal galaxies.   

In order to detect the presence of bars, DSS images of SSRS2 galaxies were visually 
inspected (by MAGM). We do not differentiate between strong and weak bars. We are 
not able to identify the totality of bars due to the saturation of the central 
parts of galaxies in DSS plates or the inability to detect bars in edge-on systems and 
for galaxies with small apparent sizes. Nevertheless, we are able to detect, at least, 
the relative differences between the SSRS2 and AGNC, since the procedure used to 
identify bars for both samples was identical.

The fraction of barred galaxies as a function of morphological type, for the SSRS2 and 
the AGNC, is displayed in Figure 9. We find that the SSRS2 shows no dependence 
between the fraction of galaxies with bars and morphology. On the other hand, the 
higher frequency of barred galaxies in the  AGNC sample is noticeable.   The barred 
fraction in the SSRS2 is 14.1\% while for the AGNC is 28.4\%.  These distributions are 
different at the 99.97\% confidence level. If we consider only the interval of 
morphological types between S0a and Sbc, where we have the majority of Seyferts, the 
result is still the same. The percentage of barred systems in the SSRS2 is smaller 
than that found in the RSA catalog by \citet{Ho97b}. This discrepancy can be partly 
explained by our inability to detect bars visually in galaxies with apparent diameters 
smaller than 0.5$\arcmin$.  

The distribution of the fraction of galaxies with bars as a function of morphology, but 
now discriminating between Seyfert 1 and Seyfert 2 galaxies is shown in Figure 10. 
The distribution is fairly noisy, and by applying a KS test we find that the 
difference between both samples is not significant.


\subsection{The Spatial Density of Seyferts}

To determine the spatial density of Seyfert galaxies in the SSRS2, we use the the 
non-parametric step-wise maximum likelihood method of \citet{eep} to measure the 
luminosity function of AGN hosts. Although in principle we could calculate the 
parametric \citet{Sch76} function, we have chosen not to do so because of the 
small sample size and the strong correlation between the different Schechter 
parameters. By calculating the step-wise maximum likelihood for both the SSRS2 
sample and the Seyfert galaxy subsample we can compare the spatial density of AGN 
hosts relative to the parent sample as a function of absolute magnitude.

Prior to the calculation of the luminosity function, galaxies with corrected radial 
velocities smaller than 500 kms$^{-1}$ or greater than 15000  kms$^{-1}$ are removed 
from the sample. The SWML is calculated for galaxies in the range 
-22.5~$\leq~M_B~\leq$~-16.0, and is shown in Figure 11 where the AGN sample
is represented as solid squares. The error bars were obtained following \citet{eep}. 
The histogram showing the number of galaxies per absolute magnitude bin is shown in 
the lower part of the figure.

In Figure 11, in addition to the AGNC non-parametric luminosity function, we show 
previous determinations by other authors converted into 
the cosmological model and magnitude system adopted in this paper: \citet{Huc92} 
represented as crosses; \citet{Koh97}, open circles; \citet{Lon00}, open triangles; 
\citet{Ulv01}, shown as solid triangles, and \citet{Gro01} as stars. The figure
also shows the SWML calculated for the SSRS2 parent sample (solid
triangles).  The AGNC luminosity function is in very good agreement with
\citet{Gro01} and \citet{Huc92}. The agreement with both \citet{Koh97} and  
\citet{Lon00} is not as good, which could be explained by the rather smaller samples 
of those two works.  The AGNC number density \citep[as well as those of][]{Huc92, 
Gro01} are systematically lower than measured by \citet{Ulv01}, probably 
reflecting the presence of LINERs in the AGN sample of the latter work.

In Figure 12 we show the ratio between the number density of AGN hosts
relative to the that of the parent sample as a function of absolute magnitude. In the 
figure we show this ratio as a differential distribution (open triangles) where we 
calculate the ratio at each magnitude bin. We also show the cumulative distribution 
(solid circles), where all galaxies in each subsample in a given magnitude
bin and brighter are added. The figure, also suggests that on average
slightly more that 1\% of galaxies down to $M_B$=-16 are AGN hosts. However, as 
brighter absolute magnitudes are probed, the proportion of AGN hosts increases, a 
trend already noted by Huchra \& Sargent (1973) and \citet{Huc92}. At $M^*$, 
corresponding to $M_B\sim$ -20.5, AGNs are found in $\sim$ 3\% of galaxies.


\section{Dependence of the AGN Phenomenon on the Environment}

In this section, we examine how the AGN phenomenon correlates with the environment 
where the host galaxy is located. An unresolved issue in the study of AGNs is 
fueling the ``central motor'' of the AGN, whereby mass in the form of gas (or stars) 
is fed into the mass concentration generating the AGN.
Among the mechanisms proposed to explain the transportation of gas from the
disk of a spiral galaxy to its nucleus, is the interaction with a close galaxy 
\citep[e.g.;][]{Nog88,Bar92}.  Several observational studies claim that AGN hosts 
show an excess of companions relative to normal galaxies 
\citep[e.g.;][]{Sta82,Dah85,Mac90,Lau94, Raf95}, while other works find no evidence 
of such an excess of companions \citep[e.g.,][]{Ulv85, Fue88, deR98}. 
Part of this controversy may be explained by the way the AGN and control samples are 
generated.  The influence of the environment over AGN phenomenon can be examined 
from scales ranging from the typical separation of close companions ($\sim$100 kpc) to
that of  groups and clusters ($\sim$ 1 Mpc).
In particular, at large scales, it is possible that even competing mechanisms could
play a complex role. In such places the AGN may be fueled from gas captured from 
close companion, during an interaction with a close companion. On the other hand, the 
central monster could be "starved" by the removal of gas from the galaxy through the 
mechanism of ram pressure stripping of the intracluster medium, as has already been 
proposed to explain anemic galaxies.

\subsection{Seyferts and Groups}

One of the first works showing that Seyfert galaxies follow the general trend of 
clustering in the same way as normal galaxies do, but avoiding rich clusters, was by 
\citet{Pet82}. This trend was subsequently confirmed by \citet{Dre85}, who found that 
the fraction of AGNs in rich clusters is $\approx$1\%, similar to the proportion of 
Seyfert galaxies in the field galaxy population. In a sample of galaxies selected in 
regions of high density of galaxies, typical of the central parts of clusters and
compact groups of galaxies, \citet{Mai94} find an excess of Seyferts when compared to 
a control sample of galaxies from regions with very low density of objects.
This result is confirmed by \citet{Foc02}, using the UZC-CG catalogue to identify
compact groups of galaxies. The proportion of galaxies hosting an AGN in compact 
groups is still not well known, and estimates range from 70\% \citep{Coz00} to 50\% 
\citep{Shi00}. In fact, the latter authors conclude, that after correcting for the
different morphological distributions of those environments, the dense galaxy 
environment in the compact groups triggers neither the AGN activity nor the nuclear 
starburst. 

The proportion of AGNs in lower density (``loose'') groups is even less well known. 
\citet{Kol89} compare the emission-line properties of galaxies in a small sample of 
groups with and without Seyferts.  They conclude that groups containing Seyferts
contain more frequently objects with strong emission lines indicative of intense 
star formation,  relative to groups without Seyfert galaxies.
In another analysis of loose groups of galaxies, \citet{Kel98} find no significant 
differences in the dynamical properties (velocity dispersions) of groups harboring 
Seyferts as compared to those without Seyferts.

We re-examine the correlation between the presence of the AGN phenomenon with
local density by identifying groups in the SSRS2. This is done using a percolation 
algorithm applied to the SSRS2 catalog. We generated a catalog of groups of galaxies 
similarly to  \citet{Mer00}, for which we examined the distribution 
of Seyfert galaxies. 
The algorithm is that described by \citet{Huc82} and \citet{Mai89}. Since we
are interested in "real physical systems", the adopted friends-of-friends algorithm, 
searches for possible group member galaxies, keeping a fixed surrounding density 
contrast ($\delta\rho/\rho$) relative to the mean density of galaxies of the
entire sample.  Groups were selected so they correspond to a density enhancement  
$\delta\rho/\rho = 80$. This level was adopted following \citet{Ram97} who showed 
that this density contrast level gives the best compromise between identifying as 
many physical loose groups as possible, including systems with high velocity 
dispersion, but minimizing contamination of the catalog by pseudo-groups as well as
groups with interlopers.  The search for companions around galaxies is carried out 
taking into account projected separations satisfying 
$$D_{12} = 2 \sin\left(\frac{\theta_{12}}{2}\right) \frac{V}{H_0}  \leq D_L$$
\noindent and with line-of-sight velocity differences,
$$V_{12} = \mid V_1 - V_2 \mid  \leq V_L$$
\noindent In the above expressions $V = (V_1 + V_2) / 2$, $V_1$ and $V_2$ are
the radial velocities of the galaxies, and $\theta_{12}$ their angular separation.  
The quantities $D_L$ and $V_L$ are search parameters scaled according to the 
expressions below in order to take into account the variation in the sampling 
of the galaxy luminosity function, $\phi(M)$, with distance
$$ D_L = D_0 R \quad ; \quad  V_L = V_0 R$$
\noindent where    
$$ R = \left[ \int_{-\infty}^{M_{12}} \Phi (M) dM \bigg/
  \int_{-\infty}^{M_{lim}}\Phi (M) dM\right]^{-1/3} $$
$$ M_{lim} = m_{lim} -25 - 5\log(V_f / H_0) $$ 
$$ M_{12} = m_{lim} -25 -5\log[(V_1 + V_2)/2 H_0] $$
\noindent $D_0$ is the selection parameter at a fixed fiducial radial velocity,  
$V_f$.  $V_L$ is scaled in the same way as $D_L$.
The adopted values for $D_0$ and $V_0$ are 0.352 h$^{-1}$Mpc and 350 km s$^{-1}$ 
respectively. The apparent magnitude, $m_{lim} = m_{B} = 15.5$, while the luminosity 
function parameters are $\phi^*=0.0137$ galaxies mag$^{-1}$ Mpc$^{-3}$,  
$M^*_{B} =-19.40$ and $\alpha=-1.08$ \citep{Mar98}.
  
The resulting group catalog contains systems with at least 4 members and mean radial 
velocities $V_g \le$ 12,000 km s$^{-1}$ (see \cite{Mer00} for more details).  
The algorithm also outputs lists of galaxies for which no companions are found 
(``isolated''), as well as binary and triple systems.  The distribution of AGNC and 
the entire SSRS2 in each of these multiplicity classes is presented in Table 2. 
An inspection of the table suggests that AGNs tend to be more common in binary 
systems, showing no preference for being located in groups with three or more 
members when compared to the control sample.

The catalogue of groups was also used to examine the distributions of velocity 
dispersions for groups containing Seyferts against the entire group sample. 
If galaxy interactions are supposed to enhance/trigger the nuclear activity, we 
should expect that the systems presenting the lowest velocity dispersions
are more likely to contain Seyfert hosts. The median values are 166
kms$^{-1}$ for groups with Seyferts and 170 km s$^{-1}$ for the entire 
sample, which are essentially identical. The same conclusion is found
by applying the KS-test to the distribution of velocity dispersions, therefore 
confirming the results of \citet{Kel98}.  
For binaries we find a very weak tendency (at the 50\% confidence level) for
Seyfert galaxies to be present in systems with higher median velocity
dispersions (102 km s$^{-1}$), relative to the entire sample (78 km s$^{-1}$).

The average group density, $\rho_g$, can be calculated by means of the
expression  $\rho_g = 3 N_g / 4 \pi r^3$ where, $N_g$ is the number of
galaxies and $r$ the mean pairwise separation of a given group.  The
distributions of $\rho_g$, show no significant difference between
Seyfert and non-Seyfert groups for binary, triple or even higher
multiplicity orders. In Figure 13 we display the median values and upper
and lower quartiles for $\rho_g$ for several group multiplicities, and
there is no evidence of the AGN phenomenon being correlated with the
group density.

\subsection{The Nearest Neighbor Distance and Local Density}

One of the suggestions of \citet{Mac90} is that close companion galaxies enhance 
the star formation rather than directly enhancing Seyfert-like activity. On the other 
hand \citet{Lau95} conclude that Seyferts have on average about twice the number of 
companions than other galaxies in their control sample.  \citet{Dul99}
also find a significant excess of large companions for Seyfert 2 galaxies but
no such evidence for Seyfert 1. On the other hand, \citet{Raf95} report no difference 
between the presence of close companions for Seyfert galaxies relative to non-active 
galaxies. We address this issue estimating the separation of the nearest neighbor, 
$S$, for galaxies in our sample.  

In this analysis we use the SSRS2 catalogue restricting it to contain objects with 
velocities smaller than 12000 km s$^{-1}$, therefore minimizing the problem of 
decreasing completeness of the sample with distance. We also neglect galaxies which 
are located at projected distances $<$ 1 Mpc from the survey boundaries. 
We further restrict the sample to galaxies with morphological types between S0a
and Sbc, the interval which contains approximately 80\% of the Seyfert galaxy
sample.  This is done to minimize the bias induced by the morphology-density relation, 
since we have a negligible fraction of Seyferts hosted by elliptical galaxies, which 
are objects generally found in high density environments. 
The distributions of the nearest neighbor, $S$, for AGNC and SSRS2 samples were 
examined using the KS test which shows that both samples are part of the same parent 
population at the 98\% confidence level. If we compare the distribution of the second 
nearest neighbor, a similar result is found. We find a trend of Seyfert 1s presenting, 
on average, closer companions than the type 2s, but the result is not statistically
significant. 
 
The nearest neighbor distance is also used to compute the maximum tidal influence, $Q$, 
that a companion may exert on the AGN host galaxy. This tidal influence is proportional 
to the companion's mass divided by the cube of the separation, $S$.  Assuming that light
traces mass, the tidal parameter for each host is $Q \propto L/S^3$.
The distribution of $Q$ values for the AGNC and SSRS2 are identical at the 86\% 
confidence level. The Seyfert types 1 and 2 are considered the same only at the 20\%
confidence level, with type 1s presenting higher values of $Q$, compared to type 2s.


\section{Summary}

In the present work we have used the magnitude-limited SSRS2 to identify galaxies 
with active galactic nuclei, and to investigate how the properties of AGN host 
galaxies relate to general population of galaxies in the parent sample. 
The main results of this work are summarized below.
\begin{itemize}

\item We find that 162/5339 (3\%) of the galaxies in the SSRS2 are
  Seyfert galaxies. This number is comparable to other works studying
  the local universe \citep[e.g.,][]{Huc92, Ho01, Gro01, Hao03}, as
  well as with the number of AGNs detected in a high redshift sample
  \citep[e.g.,][]{Sar02}. 
  
\item The ratio of Seyfert 2s to 1s is about 3:1. 

\item The majority of the AGN host galaxies in the SSRS2 are in
  systems classified morphologically betwen S0a and Sb, which are
  galaxies containing significant bulge components. This result is
  consistent with findings of \citet{Ho01}.  We find that
  $\sim$ 10\% of the Seyfert galaxies are found in hosts with indication of an 
  ongoing merger event. There is no significant difference between the
  morphological distributions of Seyfert 1 relative to Seyfert 2 hosts.

\item The AGNs are preferentially detected in high luminosity hosts
  ($M < M^*+1$). This result is similar to previous findings
  \citep[e.g.,][]{Huc92}, though in conflict with that found by \citet{Hao03} using 
  the SDSS database.  This discrepancy can be explained by the fact that the present 
  sample does not allow separating the contribution to the total luminosity due to 
  the AGN from that of the host galaxy.

\item Seyfert galaxies are twice as frequent in barred hosts than
  in galaxies without nuclear activity.

\item Using a percolation algorithm we find that Seyfert galaxies are
 more commonly found in binary systems than in groups with higher
 multiplicities or as ``isolated'' galaxies, when compared to galaxies
 in the parent sample.

\item We do not find strong correlation between the presence of an AGN with
  indicators of the strength of gravitational perturbation by
  companions, such as the projected local density, the group
  velocity dispersion, the nearest neighbor distance and the maximum tidal
  influence.

\item We find marginal evidence that Seyfert 1 galaxies have closer
  companions and are more susceptible to tidal effects than Seyfert 2
  galaxies are.

\item The results above suggest that  the large-scale environment
  does not seem to play an important role in enhancing the AGN
  activity.  On the other hand, internal characteristics such as
  morphology, luminosities, existence of bars and other asymmetries are more
  strongly correlated with the presence of the AGN.

\end{itemize}


\acknowledgments

The authors thank the referee, Dr. John Huchra, whose suggestions allowed us to 
improve the text. We also thank Alberto Rodr\'\i guez-Ardilla and Lei Hao for 
useful discussions and Caryl Gronwall, Vicki Sarajedini and John Salzer for providing 
the KISS non-parametric luminosity function.
This research has made use of NASA/IPAC Extragalactic Database (NED).  
RSM acknowledges financial support from PIBIC-CNPq scholarship, MAGM to CNPq 
grant 301366/86-1 and FAPERJ E-26/171.619/2001, and CNAW to NSF AST 95-29028 
and AST 00-71198. 



\clearpage

\begin{deluxetable}{lllrrrcrr}
\tabletypesize{\small}
\tablewidth{0pt}
\tablenum{1}
\tablecolumns{9}
\tablecaption{SSRS2 Seyfert Galaxies}
\tablehead{
  \multicolumn{2}{c} {Coordinates (J2000.0)} & 
Identification &  
        $m_{SSRS}$  & 
        Velocity  & 
        Type   & 
        Bar  & 
        $b/a$ &
        Sey \\ 
        \colhead{$\alpha$ ($^h \ ^m \ ^s$)}  & 
        \colhead{$\delta$ ($\degr \ \arcmin \ \arcsec$)}  & 
  \colhead{} & 
  \colhead{} & 
  \colhead{$km s^{-1}$} &
  \colhead{} & 
  \colhead{} & 
  \colhead{} & 
  \colhead{} 
}
\startdata 
00 03 32.1 & -10 44 41 & NGC 7808             & 14.47 &  8922 & -2 & N & 1.00 & S1 \\
00 53 29.8 & -08 46 04 & NGC 0291             & 14.64 &  5695 &  2 & Y & 0.45 & S2 \\
00 10 01.0 & -04 42 38 & MRK 0937             & 14.70 &  8851 &  3 & N & 1.00 & S1 \\
00 10 54.3 & -21 04 03 & ESO 538- G 025       & 15.02 &  7782 &  2 & N & 0.22 & S2 \\
00 18 35.9 & -07 02 56 & GSC 4670 00946       & 14.36 &  5620 &  0 & N & 1.00 & S2 \\
00 19 44.0 & -14 07 18 & IC 0009              & 15.42 & 12622 &  0 & N & 0.75 & S2 \\
00 26 00.0 & -02 55 07 & MCG -01-02-013       & 15.49 & 20282 & 15 & N & 0.72 & S2 \\
00 30 53.8 & -09 12 07 & MCG -02-02-035       & 15.28 &  5939 & -2 & N & 0.50 & S2 \\
00 34 13.8 & -21 26 21 & ESO 540- G 001       & 13.96 &  8030 &  4 & Y & 0.69 & S1 \\
00 35 48.8 & -13 36 38 & NGC 0166             & 15.18 &  6020 &  0 & N & 0.33 & S2 \\
00 38 18.4 & -14 50 07 & MCG -03-02-027       & 15.40 & 10971 &  3 & N & 0.67 & S2 \\
00 41 11.7 & -21 07 54 & ESO 540- G 014       & 15.38 &  1653 & -5 & N & 0.50 & S2 \\
00 42 52.8 & -23 32 29 & NGC 0235A            & 14.18 &  6589 & 15 & N & 0.44 & S2 \\
00 54 54.5 & -32 01 54 & ESO 411- G 029       & 15.04 &  9622 &  0 & N & 0.38 & S2 \\
00 55 02.6 & -19 00 18 & ESO 541- G 001       & 13.61 &  6366 &  3 & Y & 0.62 & S2 \\
00 58 22.2 & -36 39 37 & ESO 351- G 025       & 15.12 & 10415 &  2 & N & 0.58 & S2 \\
01 02 17.4 & -19 40 09 & ESO 541-IG 012 NED01 & 15.22 & 16905 & 15 & N & 0.85 & S2 \\
01 08 47.6 & -15 50 34 & IC 0078              & 14.46 & 12066 &  1 & N & 0.41 & S2 \\
01 11 27.5 & -38 05 01 & NGC 0424             & 13.90 &  3496 &  0 & N & 0.50 & S1 \\
01 12 19.2 & -32 03 43 & NGC 0427             & 14.87 & 10011 & -2 & Y & 0.62 & S1 \\
01 14 07.0 & -32 39 03 & IC 1657              & 13.12 &  3564 &  3 & N & 0.25 & S2 \\
01 19 25.0 & -15 41 07 & MCG -03-04-046       & 15.26 & 15234 &  4 & Y & 1.00 & S2 \\
01 23 54.4 & -35 03 56 & NGC 0526A            & 14.66 &  5725 & 15 & N & 0.53 & S2 \\
01 31 50.4 & -33 08 10 & ESO 353- G 009       & 14.07 &  4970 &  2 & Y & 1.00 & S2 \\
01 39 24.8 & -09 24 04 & NGC 0640             & 15.45 &  7490 &  2 & N & 0.67 & S2 \\
01 43 37.6 & -33 42 20 & ESO 353- G 038       & 14.85 &  8884 &  5 & N & 0.36 & S2 \\
01 51 41.7 & -36 11 16 & ESO 354- G 004       & 15.08 & 10045 &  3 & N & 0.93 & S1 \\
01 59 51.3 & -06 50 25 & IC 0184              & 14.87 &  5382 &  1 & N & 0.50 & S2 \\
02 01 06.5 & -06 48 56 & NGC 0788             & 13.50 &  4078 & -5 & N & 0.77 & S2 \\
02 06 53.1 & -36 27 08 & NGC 0824             & 14.05 &  5836 &  2 & Y & 0.95 & S2 \\
02 09 20.9 & -10 08 00 & NGC 0833             & 13.64 &  3867 &  3 & N & 0.47 & S2 \\
02 10 11.4 & -09 03 36 & GSC 5281 00378       & 15.23 & 12491 &  3 & N & 0.50 & S2 \\
02 24 40.6 & -19 08 30 & ESO 545- G 013       & 13.59 & 10253 &  3 & N & 0.73 & S1 \\
02 30 05.5 & -08 59 53 & MCG -02-07-024       & 15.25 &  4859 & -2 & N & 0.86 & S1 \\
02 31 51.0 & -36 40 16 & IC 1816              & 13.66 &  5215 &  3 & N & 0.93 & S2 \\
02 34 37.8 & -08 47 15 & NGC 0985             & 15.00 & 19150 & 15 & N & 1.00 & S1 \\
02 35 13.5 & -29 36 17 & ESO 416- G 002       & 15.06 & 17700 &  1 & N & 0.50 & S2 \\
02 38 45.2 & -30 48 24 & ESO 416- G 005       & 15.34 & 18587 &  5 & N & 0.80 & S1 \\
02 41 04.8 & -08 15 21 & NGC 1052             & 12.08 &  1519 & -3 & N & 0.70 & S2 \\
02 41 38.7 & -28 10 17 & IC 1833              & 14.28 &  4952 & -2 & N & 0.58 & S2 \\
02 43 07.8 & -08 46 26 & NGC 1071             & 15.41 & 11302 &  3 & N & 0.45 & S2 \\
02 44 02.9 & -26 11 11 & ESO 479- G 030       & 15.29 & 10504 &  2 & N & 0.24 & S2 \\
02 46 19.0 & -30 16 29 & NGC 1097             & 10.23 &  1195 &  3 & N & 0.62 & S1 \\
02 49 33.8 & -38 46 12 & ESO 299- G 020       & 13.96 &  5008 &  1 & N & 0.45 & S2 \\
02 51 40.3 & -16 39 04 & NGC 1125             & 13.87 &  3303 &  0 & Y & 0.50 & S2 \\
02 56 09.8 & -13 41 07 & MCG -02-08-031       & 15.38 &  6935 &  3 & N & 0.50 & S2 \\
02 56 21.5 & -32 11 06 & ESO 417- G 006       & 14.34 &  4947 & -2 & N & 0.83 & S2 \\
02 57 49.6 & -10 10 07 & MCG -02-08-033       & 14.02 &  4516 &  3 & N & 0.30 & S2 \\
03 00 04.3 & -10 49 28 & MCG -02-08-038       & 15.27 &  9770 &  0 & Y & 0.50 & S1 \\
03 00 30.7 & -11 24 07 & MCG -02-08-039       & 14.57 &  8989 &  3 & Y & 0.69 & S2 \\
03 02 13.2 & -23 35 20 & GSC 6438 00439       & 14.93 & 10514 & -2 & N & 0.39 & S1 \\
03 08 10.8 & -22 57 39 & NGC 1229             & 14.82 & 10676 &  1 & N & 0.69 & S2 \\
03 11 14.7 & -08 55 20 & NGC 1241             & 13.26 &  4036 &  3 & Y & 0.61 & S2 \\
03 24 48.7 & -03 02 32 & NGC 1320             & 13.67 &  2698 &  2 & N & 0.32 & S2 \\
03 25 04.9 & -12 18 07 & MCG -02-09-040       & 14.93 &  4400 & -2 & N & 0.33 & S2 \\
03 25 25.3 & -06 08 39 & MRK 0609             & 15.06 & 10236 & -5 & N & 0.75 & S2 \\
03 30 40.9 & -03 08 16 & MRK 0612             & 15.10 &  6195 &  3 & Y & 0.62 & S2 \\
03 33 39.7 & -05 05 22 & NGC 1358             & 13.30 &  4013 &  0 & Y & 0.77 & S2 \\
03 33 36.4 & -36 08 25 & NGC 1365             & 10.34 &  1659 &  3 & Y & 0.71 & S1 \\
03 36 46.4 & -35 59 58 & NGC 1386             & 12.64 &   924 & -2 & N & 0.38 & S2 \\
03 37 03.2 & -25 14 57 & ESO 482- G 014       & 15.39 & 12889 &  1 & N & 0.40 & S2 \\
03 42 03.2 & -21 14 50 & ESO 548- G 081       & 12.92 &  4341 &  1 & N & 0.87 & S1 \\
03 43 26.5 & -31 44 38 & GSC 7028 00092       & 14.90 &  9574 &  3 & N & 0.64 & S2 \\
03 45 12.5 & -39 34 30 & GSC 7569 01468       & 15.48 & 12884 & -5 & N & 0.96 & S1 \\
03 55 09.3 & -17 28 10 & ESO 549- G 036       & 14.64 &  8501 &  4 & Y & 0.75 & S2 \\
03 57 38.2 & -19 12 59 & NGC 1489             & 14.36 & 11421 &  3 & Y & 0.44 & S2 \\
04 02 25.7 & -18 02 51 & ESO 549- G 049       & 14.36 &  7872 &  3 & N & 0.75 & S2 \\
04 02 46.1 & -21 07 09 & ESO 549- G 050       & 14.17 &  7532 &  3 & N & 0.67 & S2 \\
04 04 27.5 & -10 10 07 & MCG -02-11-014       & 15.41 &  9267 &  2 & Y & 0.62 & S2 \\
04 13 49.7 & -32 00 25 & ESO 420- G 013       & 13.52 &  3594 &  0 & N & 0.90 & S2 \\
04 39 37.3 & -37 05 08 & GSC 7045 01444       & 15.22 & 12292 &  3 & N & 0.48 & S2 \\
04 40 59.6 & -37 34 11 & ESO 304- G 011       & 15.34 & 12392 &  3 & Y & 0.45 & S2 \\
09 42 33.3 & -03 41 55 & NGC 2974             & 12.30 &  1908 &  1 & N & 0.57 & S2 \\
09 50 56.5 & -04 59 07 & MCG -01-25-049       & 14.46 &  6628 &  1 & Y & 0.42 & S2 \\
10 02 00.1 & -08 09 42 & GSC 5476 01170       & 15.22 &  4569 & 15 & N & 0.90 & S1 \\
10 18 03.3 & -03 47 41 & GSC 4907 01728       & 15.36 & 11947 &  2 & N & 1.00 & S2 \\
10 20 57.8 & -04 57 03 & GSC 4908 01685       & 15.25 & 11889 &  1 & Y & 1.00 & S2 \\
10 25 50.2 & -11 26 07 & MCG -02-27-003       & 15.00 & 11858 &  0 & N & 1.00 & S2 \\
10 26 51.9 & -03 27 52 & IC 0614              & 15.39 & 10194 &  0 & N & 0.71 & S2 \\
10 35 27.3 & -14 07 07 & MCG -02-27-009       & 14.33 &  4577 &  2 & N & 0.26 & S2 \\
10 38 20.6 & -10 07 04 & GSC 5495 00478       & 15.41 &  8710 &  2 & Y & 0.50 & S1 \\
10 39 46.3 & -05 29 00 & MCG -01-27-031       & 14.07 &  6194 &  1 & Y & 0.75 & S1 \\
10 42 18.7 & -17 38 07 & MCG -03-27-026       & 14.80 &  6191 & -2 & N & 0.29 & S2 \\
10 57 53.6 & -04 54 18 & IC 0657              & 15.44 &  9697 &  1 & Y & 0.42 & S2 \\
11 00 45.4 & -06 34 43 & GSC 4927 01023       & 15.36 &  8997 &  3 & N & 0.24 & S2 \\
11 23 32.2 & -08 39 30 & NGC 3660             & 13.06 &  3678 &  4 & Y & 0.87 & S2 \\
11 24 56.3 & -03 48 41 & GSC 4926 01061       & 15.04 &  6788 &  2 & N & 0.40 & S2 \\
11 25 02.1 & -05 04 07 & MCG -01-29-018       & 14.53 &  8518 &  2 & N & 0.90 & S2 \\
11 34 23.3 & -09 40 07 & MCG -01-30-005       & 15.12 &  6419 & -3 & N & 1.00 & S2 \\
11 43 18.6 & -12 52 42 & NGC 3831             & 13.60 &  5254 &  1 & N & 0.22 & S2 \\
11 45 11.6 & -09 18 51 & NGC 3858             & 13.94 &  5729 &  0 & Y & 0.67 & S2 \\
11 45 40.5 & -18 27 16 & GSC 6096 00769       & 14.70 &  9877 &  2 & N & 0.55 & S1 \\
11 52 38.2 & -05 12 07 & MCG -01-30-041       & 14.70 &  5749 &  0 & Y & 0.50 & S2 \\
12 00 43.3 & -20 50 01 & GSC 6100 00370       & 15.46 & 18639 & 15 & N & 0.76 & S1 \\
12 16 60.0 & -26 12 36 & ESO 505-IG 031       & 15.49 & 11840 & 15 & Y & 0.88 & S2 \\
12 39 39.4 & -05 20 39 & NGC 4593             & 12.21 &  2698 &  2 & Y & 0.83 & S1 \\
12 39 59.4 & -11 37 23 & NGC 4594             &  9.50 &   994 &  0 & N & 0.62 & S2 \\
12 42 25.3 & -06 58 16 & NGC 4628             & 14.50 &  2828 &  1 & N & 0.23 & S2 \\
12 51 32.4 & -14 13 17 & MCG -02-33-030       & 14.23 &  4298 &  1 & N & 0.43 & S1 \\
12 52 12.4 & -13 24 54 & NGC 4748             & 14.27 &  4426 & 15 & N & 0.86 & S1 \\
12 52 36.4 & -21 54 39 & ESO 575-IG 016 NED02 & 14.87 &  6865 & 15 & N & 0.70 & S2 \\
12 56 36.3 & -06 49 03 & NGC 4813             & 14.15 &  1373 &  1 & N & 0.40 & S2 \\
13 00 52.9 & -13 26 59 & NGC 4897             & 13.23 &  2546 &  3 & Y & 0.80 & S2 \\
13 04 14.3 & -10 20 23 & NGC 4939             & 11.99 &  3117 &  3 & N & 0.38 & S2 \\
13 04 13.1 & -05 33 06 & NGC 4941             & 12.43 &  1108 &  1 & N & 0.48 & S2 \\
13 06 56.5 & -23 55 01 & IC 4180              & 13.88 &  2972 &  0 & Y & 0.77 & S2 \\
13 07 05.9 & -23 40 39 & NGC 4968             & 13.92 &  2930 &  0 & N & 0.47 & S2 \\
13 10 17.3 & -07 27 07 & MCG -01-34-008       & 14.54 &  6713 &  3 & N & 0.86 & S2 \\
13 17 03.4 & -02 15 41 & IC 4218              & 14.90 &  5808 &  1 & N & 0.27 & S1 \\
13 19 31.7 & -12 39 26 & NGC 5077             & 12.41 &  2817 & -5 & N & 0.50 & S2 \\
13 20 06.4 & -17 07 06 & MCG -03-34-049       & 15.14 &  6713 &  0 & N & 0.83 & S1 \\
13 22 24.5 & -16 43 07 & MCG -03-34-064       & 14.64 &  6393 & -2 & N & 0.80 & S2 \\
13 24 35.2 & -19 45 11 & GSC 6128 00106       & 14.97 &  5284 &  3 & N & 0.77 & S2 \\
13 27 12.7 & -24 51 41 & GSC 6717 00254       & 15.42 & 12131 &  2 & N & 0.79 & S1 \\
13 30 41.7 & -21 39 47 & GSC 6133 00078       & 15.49 &  7292 &  0 & Y & 0.73 & S2 \\
13 31 13.9 & -25 24 10 & ESO 509- G 038       & 14.74 &  7786 & 15 & N & 0.42 & S1 \\
13 32 39.1 & -10 28 53 & MCG -02-35-001       & 14.82 &  6638 &  2 & Y & 0.67 & S1 \\
13 33 08.3 & -23 32 37 & GSC 6714 00714       & 14.50 & 10342 &  3 & N & 0.81 & S2 \\
13 34 39.6 & -23 26 48 & ESO 509-IG 066       & 14.70 & 10031 & 15 & N & 0.77 & S2 \\
13 37 35.1 & -21 12 58 & GSC 6134 00646       & 15.26 & 15703 &  3 & N & 0.49 & S2 \\
13 37 50.0 & -23 59 41 & ESO 509-IG 083 NED01 & 14.70 &  9044 & 15 & N & 0.63 & S2 \\
13 40 19.7 & -23 51 29 & NGC 5260             & 13.42 &  6539 &  2 & Y & 0.88 & S2 \\
13 46 20.1 & -03 25 42 & MCG +00-35-020 NED02 & 15.02 &  6948 &  2 & N & 0.38 & S2 \\
13 58 59.7 & -20 02 44 & GSC 6144 00913       & 15.50 & 11797 & -5 & N & 1.00 & S2 \\
14 03 26.1 & -06 01 51 & NGC 5427             & 12.50 &  2618 &  3 & N & 1.00 & S2 \\
14 13 14.9 & -03 12 27 & NGC 5506             & 13.37 &  1753 &  3 & N & 0.57 & S2 \\
14 26 12.3 & -11 54 07 & MCG -02-37-004       & 15.48 & 12480 &  3 & Y & 0.80 & S2 \\
14 33 43.5 & -14 37 11 & NGC 5664             & 14.52 &  4537 &  2 & N & 0.42 & S2 \\
14 42 23.9 & -17 15 10 & NGC 5728             & 12.81 &  2793 &  4 & Y & 0.58 & S2 \\
14 59 24.8 & -16 41 36 & NGC 5793             & 14.17 &  3522 &  2 & N & 0.35 & S2 \\
15 33 20.7 & -08 42 07 & MCG -01-40-001       & 14.64 &  6888 &  4 & Y & 0.38 & S2 \\
21 07 59.9 & -29 50 10 & GSC 6936 00748       & 15.10 &  5750 &  1 & N & 0.90 & S2 \\
21 27 02.8 & -22 59 32 & ESO 530- G 047       & 14.51 &  9761 &  2 & N & 0.59 & S2 \\
21 48 19.5 & -34 57 06 & NGC 7130             & 13.33 &  4829 & 15 & N & 0.94 & S2 \\
21 56 56.6 & -11 39 31 & GSC 5797 01005       & 15.37 & 17650 &  1 & N & 0.54 & S1 \\
22 00 21.6 & -13 08 49 & IC 1417              & 14.36 &  5446 &  3 & N & 0.29 & S2 \\
22 01 17.0 & -37 46 07 & MCG -06-48-013       & 14.82 & 10005 & -2 & N & 0.75 & S2 \\
22 02 01.7 & -31 52 18 & NGC 7172             & 12.95 &  2593 &  1 & N & 0.52 & S2 \\
22 08 28.3 & -34 06 23 & ESO 404- G 032       & 14.33 &  4431 &  4 & N & 0.20 & S2 \\
22 09 07.7 & -27 48 34 & NGC 7214             & 13.05 &  6832 &  4 & N & 0.74 & S1 \\
22 14 42.0 & -38 48 23 & ESO 344- G 016       & 15.08 & 11906 &  2 & Y & 1.00 & S1 \\
22 34 49.8 & -25 40 37 & ESO 533- G 050       & 14.40 &  7928 &  0 & Y & 0.89 & S2 \\
22 35 45.9 & -26 03 01 & NGC 7314             & 11.88 &  1430 &  4 & N & 0.42 & S2 \\
22 36 46.5 & -12 32 43 & MCG -02-57-023       & 14.82 &  7169 &  4 & N & 0.30 & S1 \\
22 36 55.9 & -22 13 15 & ESO 602- G 031       & 14.88 & 10101 &  3 & N & 0.50 & S1 \\
22 47 47.6 & -11 48 59 & NGC 7378             & 13.64 &  2580 &  2 & Y & 0.64 & S2 \\
22 49 37.1 & -19 16 07 & MCG -03-58-007       & 14.84 &  9488 &  3 & Y & 0.80 & S2 \\
22 54 15.7 & -37 04 59 & ESO 406- G 018       & 15.29 & 17215 & 15 & N & 0.54 & S2 \\
22 55 01.0 & -39 39 41 & NGC 7410             & 11.61 &  1751 &  1 & Y & 0.29 & S2 \\
22 59 01.4 & -25 31 42 & ESO 535- G 001       & 13.84 &  9015 &  2 & N & 0.94 & S2 \\
23 00 47.8 & -12 55 07 & NGC 7450             & 14.00 &  3134 &  1 & Y & 1.00 & S1 \\
23 03 11.1 & -08 59 21 & IC 1464              & 14.90 &  7247 & 15 & N & 0.67 & S2 \\
23 04 43.5 & -08 41 09 & MCG -02-58-022       & 15.00 & 14185 &  1 & N & 1.00 & S1 \\
23 16 37.4 & -02 19 50 & NGC 7566             & 14.27 &  7972 &  1 & Y & 0.54 & S2 \\
23 18 22.5 & -04 24 59 & NGC 7592             & 14.50 &  7328 & 15 & N & 0.80 & S2 \\
23 25 24.2 & -38 26 51 & GSC 8013 01279       & 14.89 & 10761 &  5 & N & 1.00 & S1 \\
23 30 32.3 & -02 27 45 & MCG -01-59-027       & 15.14 & 10006 &  0 & Y & 0.68 & S1 \\
23 30 47.7 & -13 29 08 & IC 1495              & 14.14 &  6391 &  2 & Y & 0.77 & S2 \\
23 42 05.2 & -39 13 00 & GSC 8014 00236       & 15.42 & 12828 &  3 & N & 0.71 & S1 \\
23 53 19.8 & -30 09 03 & ESO 471-IG 037 NED02 & 15.24 & 14660 & 15 & N & 0.65 & S2 \\
23 57 28.0 & -30 27 41 & MCG -05-01-013       & 14.97 &  9080 &  3 & N & 0.56 & S1 \\
23 59 10.7 & -04 07 37 & IC 1524              & 13.94 &  5662 &  2 & Y & 0.62 & S1 \\

\enddata
\end{deluxetable}


\begin{deluxetable}{ccccc}
\tabletypesize{\small}
\tablewidth{0pt}
\tablenum{2}
\tablecolumns{5}
\tablecaption{SSRS2 and Seyfert galaxies versus environment assignement}
\tablehead{
                 &  
        Groups   & 
        Triplets & 
        Binaries & 
        Isolated 
}
\startdata 
Seyferts & 21\%  & 7\%  & 28\%  & 44\% \\
SSRS2    & 24\%  & 8\%  & 18\%  & 50\% \\

\enddata
\end{deluxetable}


\clearpage

\begin{figure}
\includegraphics{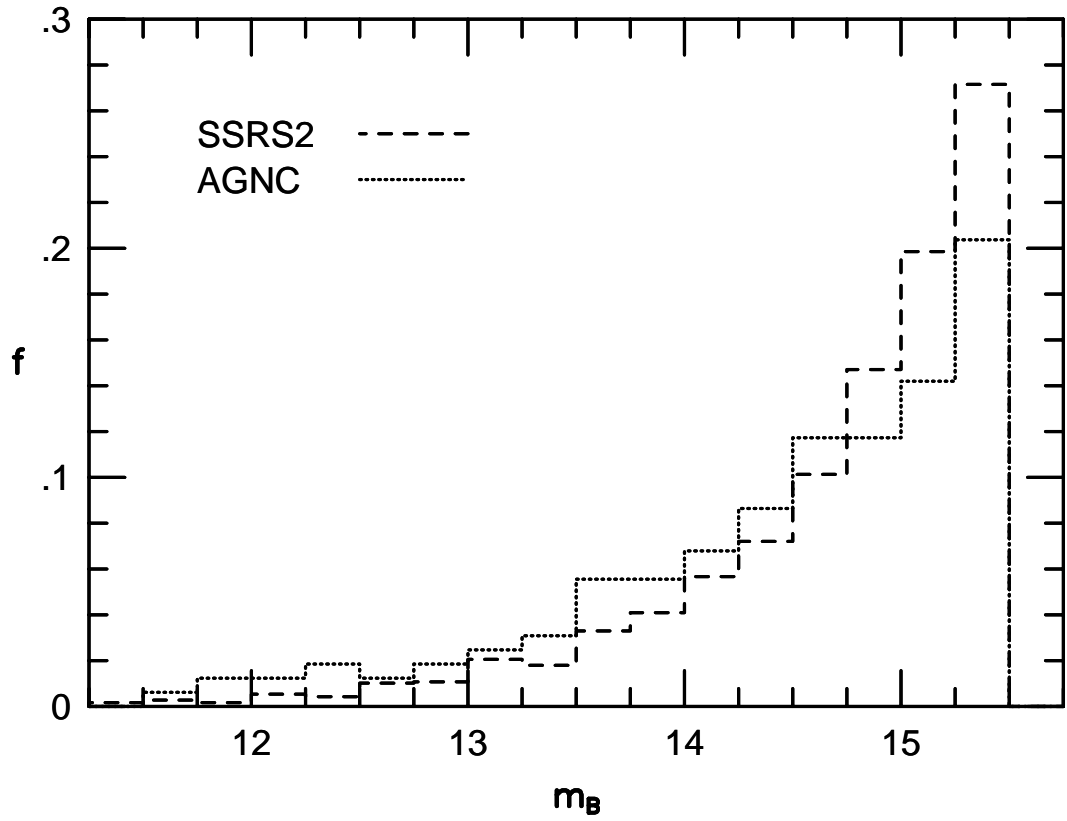}
\vspace{3cm}
\caption {Apparent magnitude distribution of galaxies in the SSRS2
(dashed line) and AGNC (solid line) samples. } 
\label {f1} 
\end{figure}

\clearpage

\begin{figure}
\includegraphics{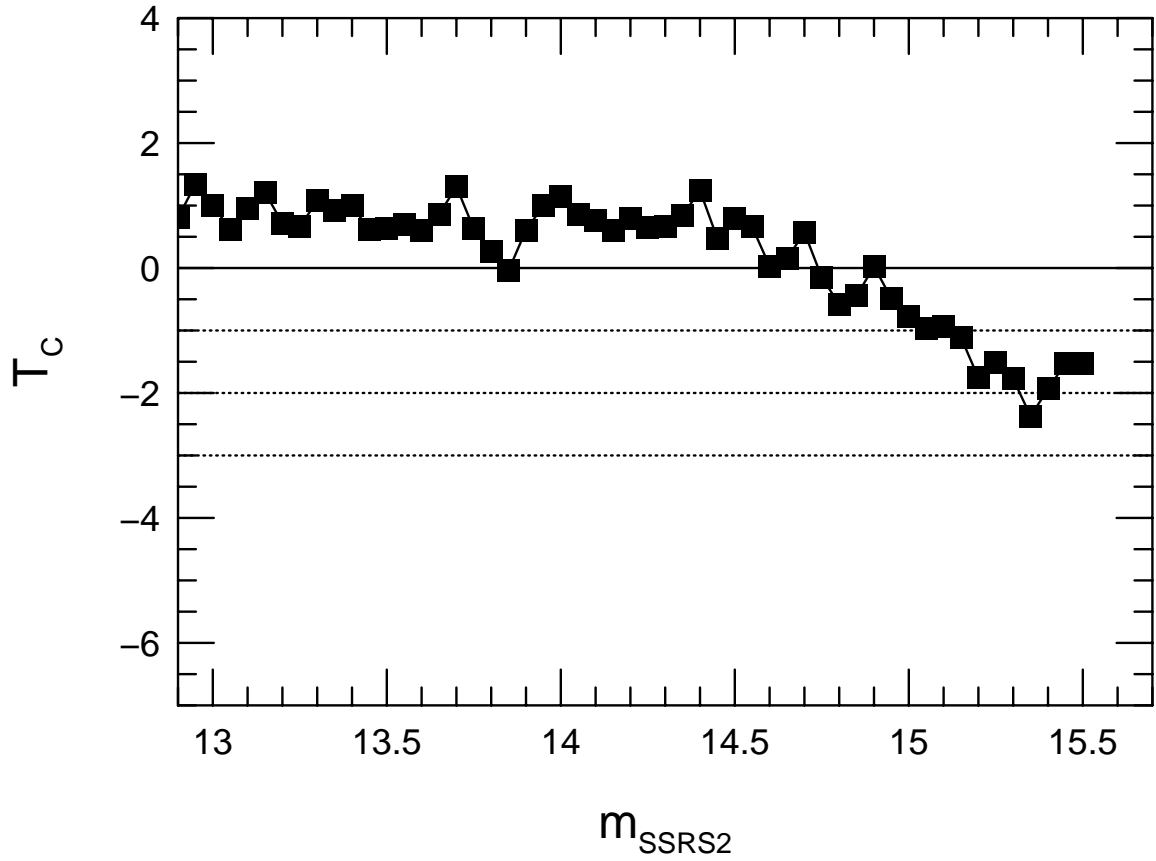}
\vspace{4cm}
\caption {Completeness test of \citet{Rau01} which shows the 
distribution of SSRS2 AGNs at different limiting magnitudes (represented as squares) 
and the different confidence levels where the catalog completeness may be rejected in 
number of standard deviations (0\%, 68\%, 90 \% and 95\%), represented by the 
horizontal lines. The sample completeness cannot be rejected at the 95\% 
confidence level at the limiting apparent magnitude.}
\label {f2} 
\end{figure}

\clearpage

\begin{figure}
\includegraphics{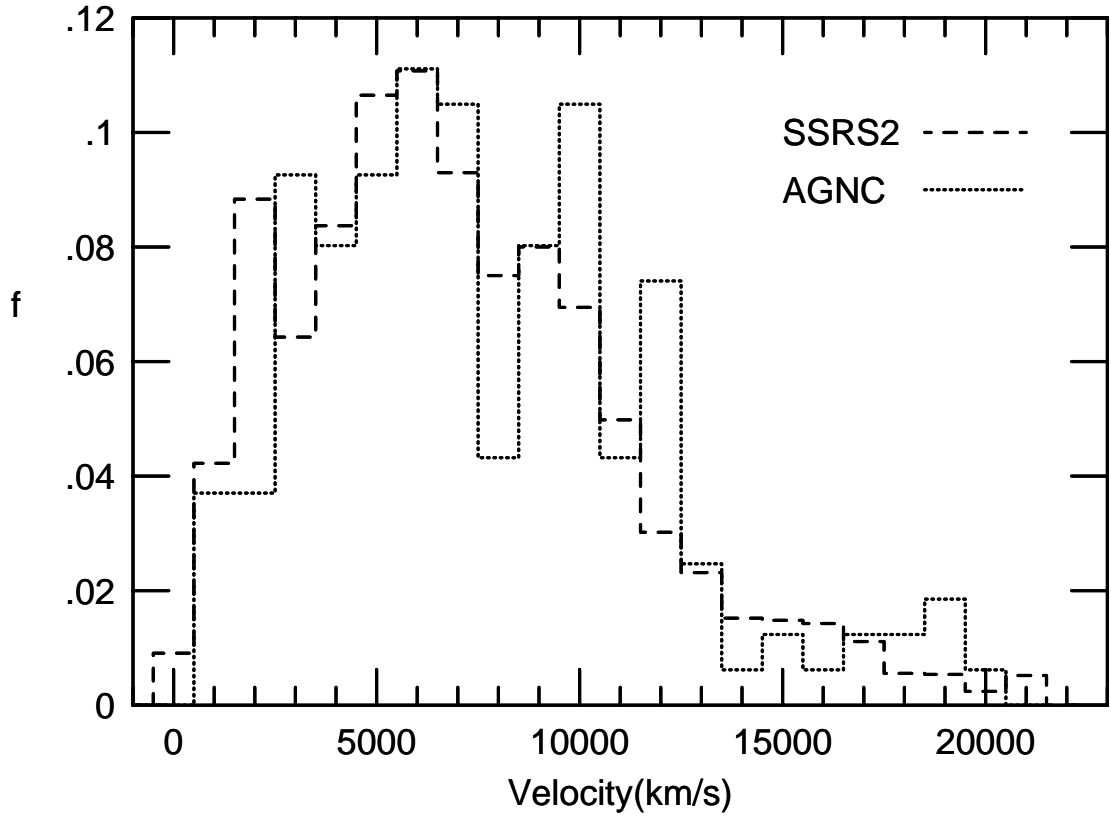}
\vspace{3cm}
\caption {Radial velocity distribution of galaxies in the AGNC (solid line) and 
SSRS2 (dashed line). }
\label {f3} 
\end{figure}

\clearpage

\begin{figure}
\includegraphics{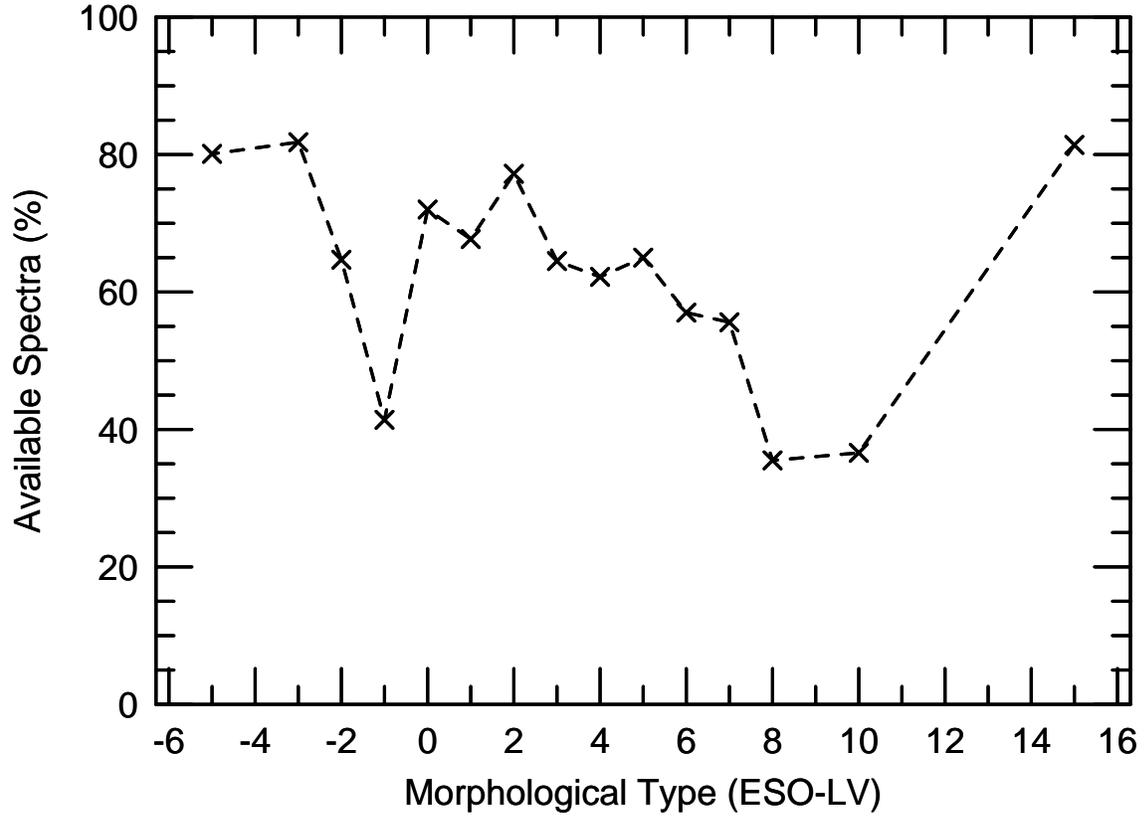}
\vspace{3cm}
\caption {Percentage of SSRS2 galaxies which have spectra 
available in our database as a function of galaxy morphology. The numbers 
correspond to the following classification: E=-5, E/S0=-3, S0=-2, S0a=0, Sa=1, 
Sab=2, Sb=3, Sbc=4, S...=5, Sc=6, Sc/Irr=7, Sd=8, Irr=10, merger=15.}
\label {f4} 
\end{figure}

\clearpage

\begin{figure}
\includegraphics{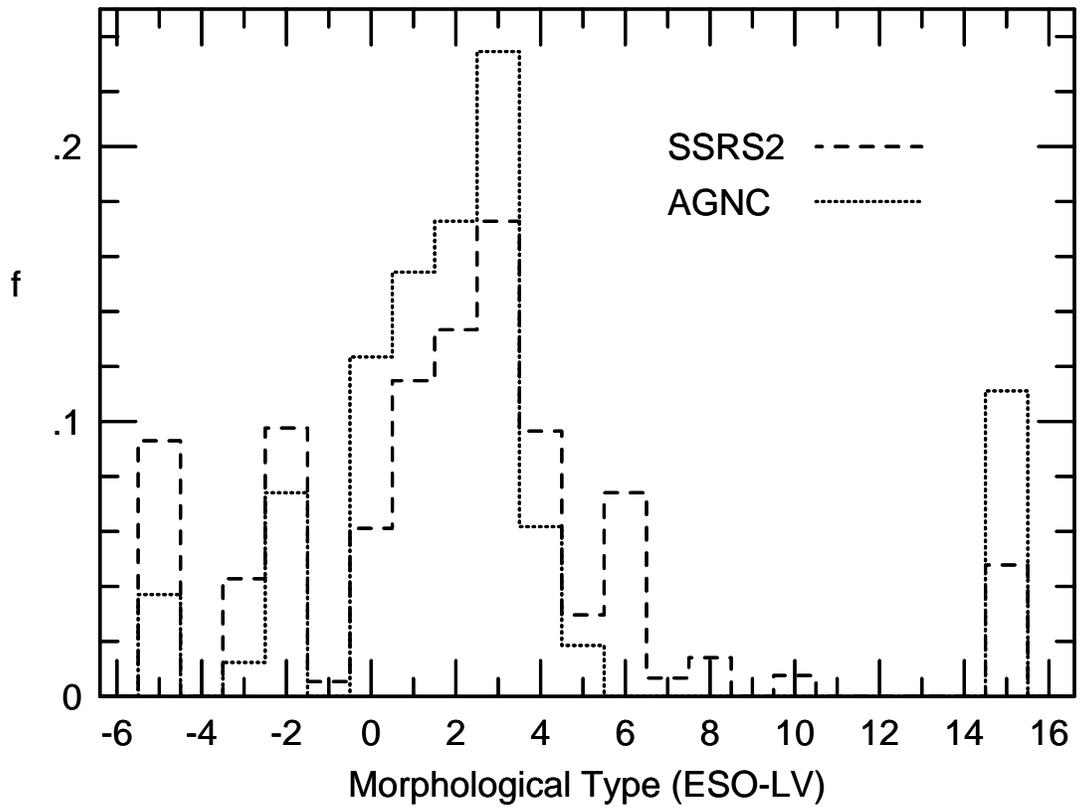}
\vspace{3cm}
\caption {Fraction of galaxies in the AGNC (solid line) and SSRS2
(dashed line) as a function of morphological type distribution.  The coding used for 
the morphological types is the same as in Fig. 4.}
\label {f5} 
\end{figure}

\clearpage

\begin{figure}
\includegraphics{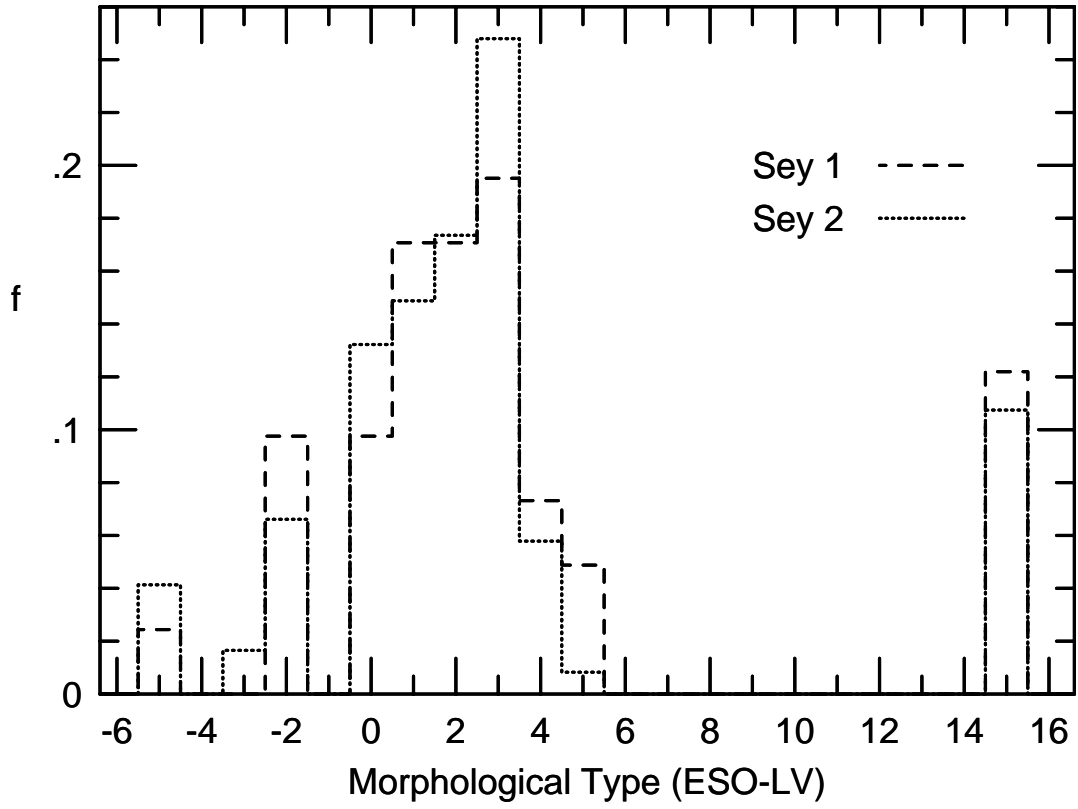}
\vspace{3cm}
\caption {Morphological distribution of Seyfert 1 and 2. The numbers
for morphological types are the same as in Fig 4. }
\label {f6} 
\end{figure}

\clearpage

\begin{figure}
\includegraphics{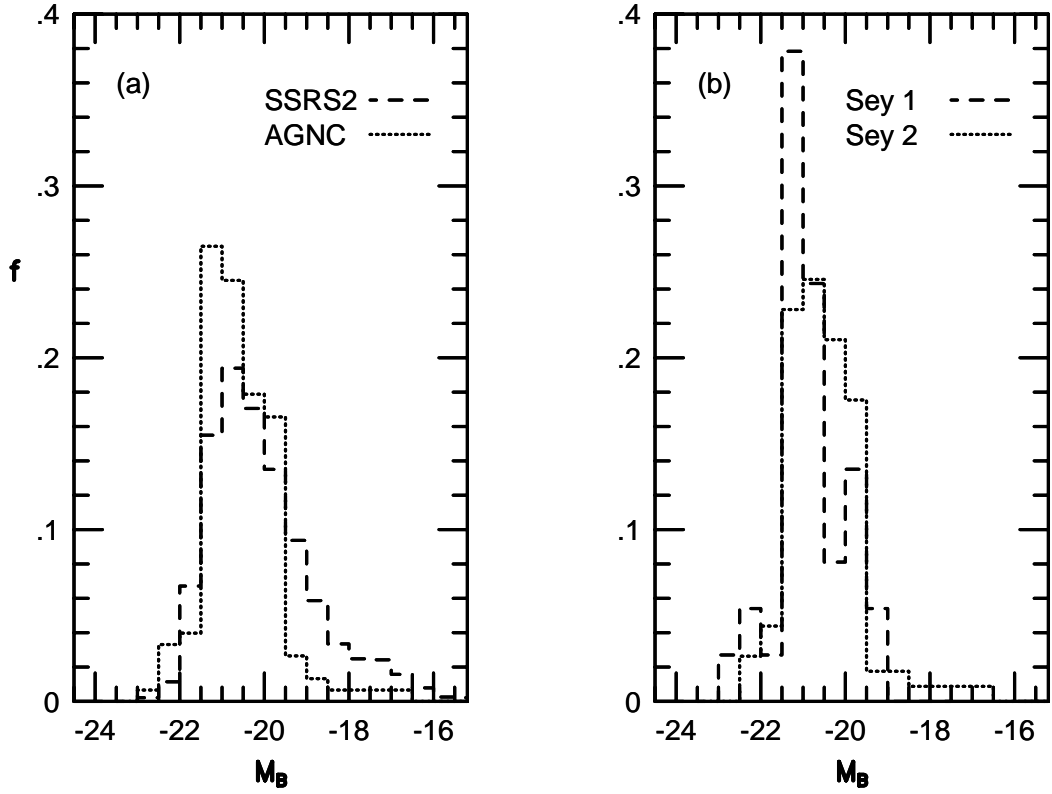}
\vspace{3cm}
\caption {Fraction of galaxies as a function of absolute magnitude 
for the AGNC (solid line) and SSRS2 (dashed line) in panel (a), while panel (b) shows 
the same distribution for the AGNC sample divided into  Seyfert 2 (solid line) and 
Seyfert 1 (dashed line) galaxies.}
\label {f7} 
\end{figure}

\clearpage

\begin{figure}
\includegraphics{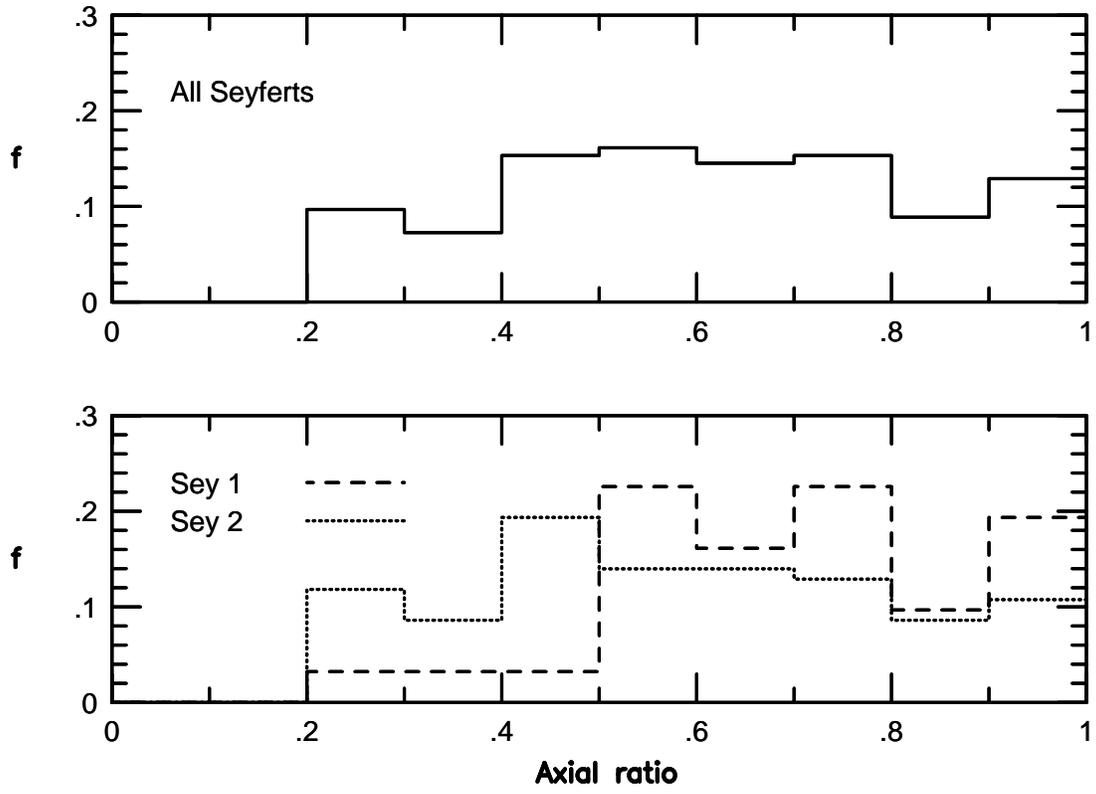}
\vspace{3cm}
\caption {Distribution of the apparent axial ratios of AGNC hosts 
(panel a) and for Seyferts 1 and 2 (panel b).}
\label {f8} 
\end{figure}

\clearpage

\begin{figure}
\includegraphics{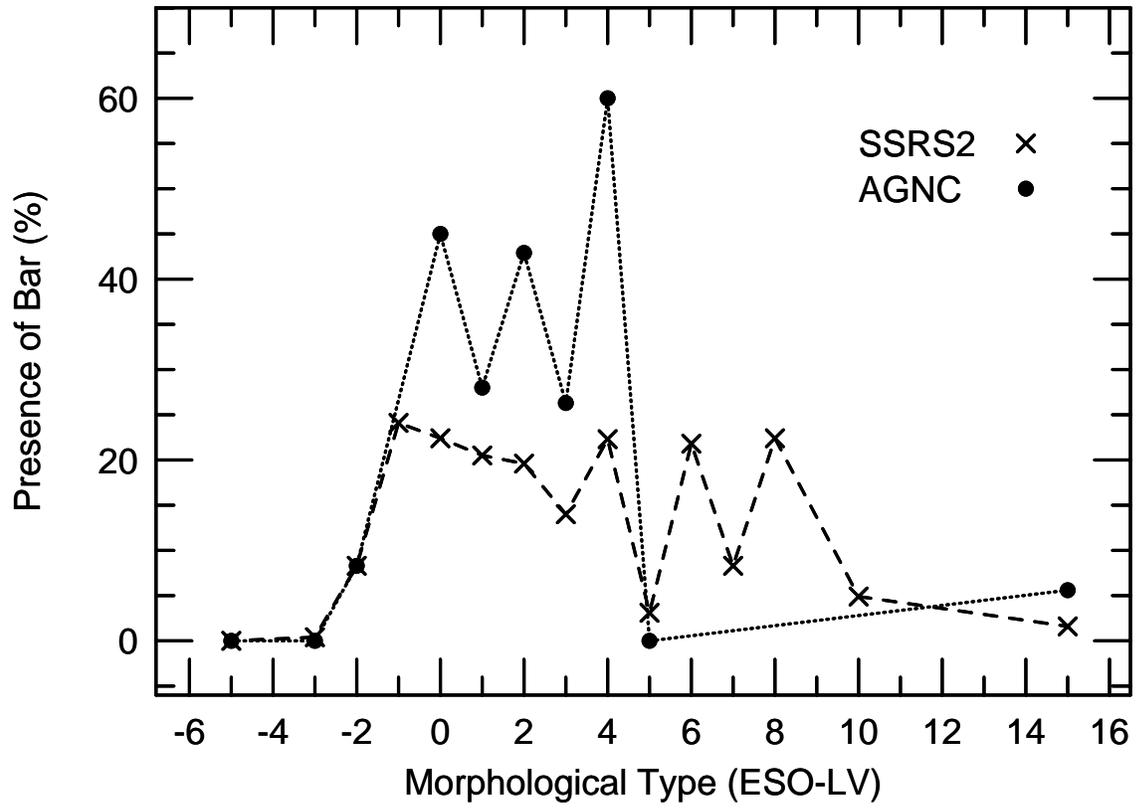}
\vspace{3cm}
\caption {Percentage of galaxies presenting bars in the AGNC (solid 
line) and  SSRS2 (dashed line) as a function of morpholgical type, using the same 
coding as in Fig. 4.}
\label {f9} 
\end{figure}

\clearpage

\begin{figure}
\includegraphics{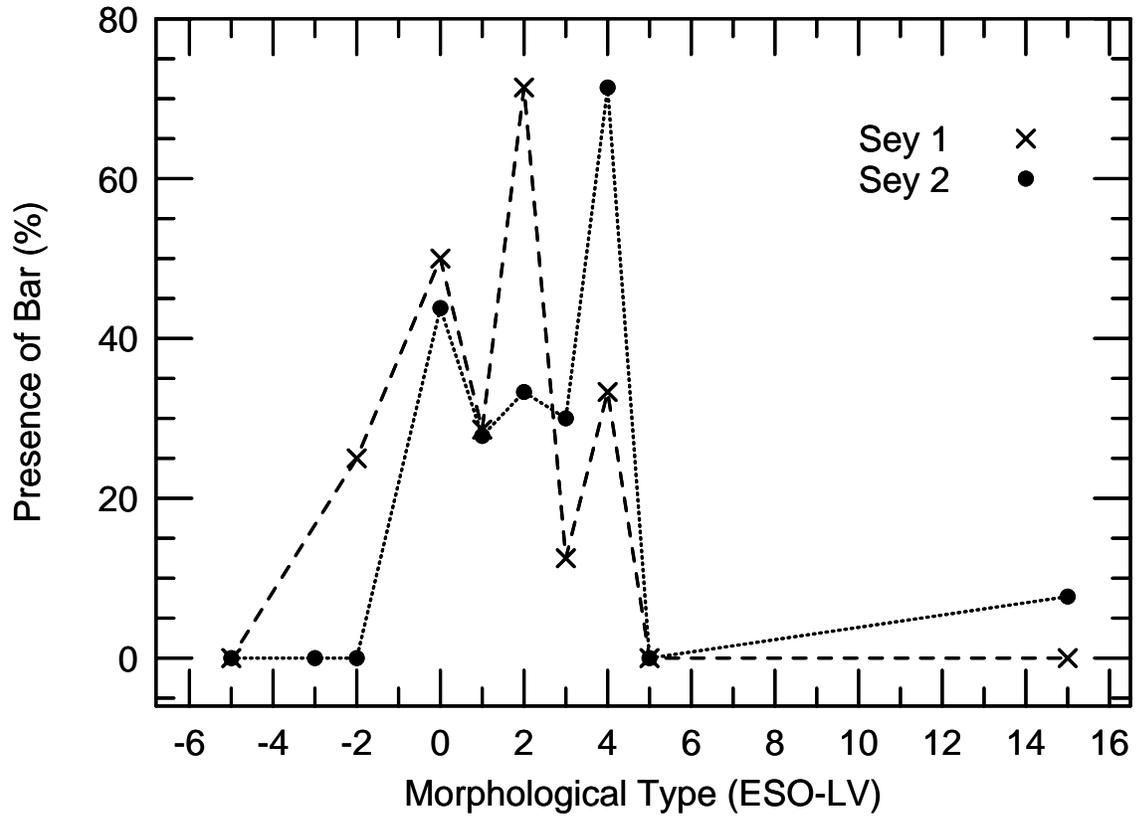}
\vspace{3cm}
\caption {Percentage of bars of Seyferts 1 and 2. The numbers for 
morphological types are the same as in Fig. 4.}
\label {f10} 
\end{figure}

\clearpage

\begin{figure}
\includegraphics{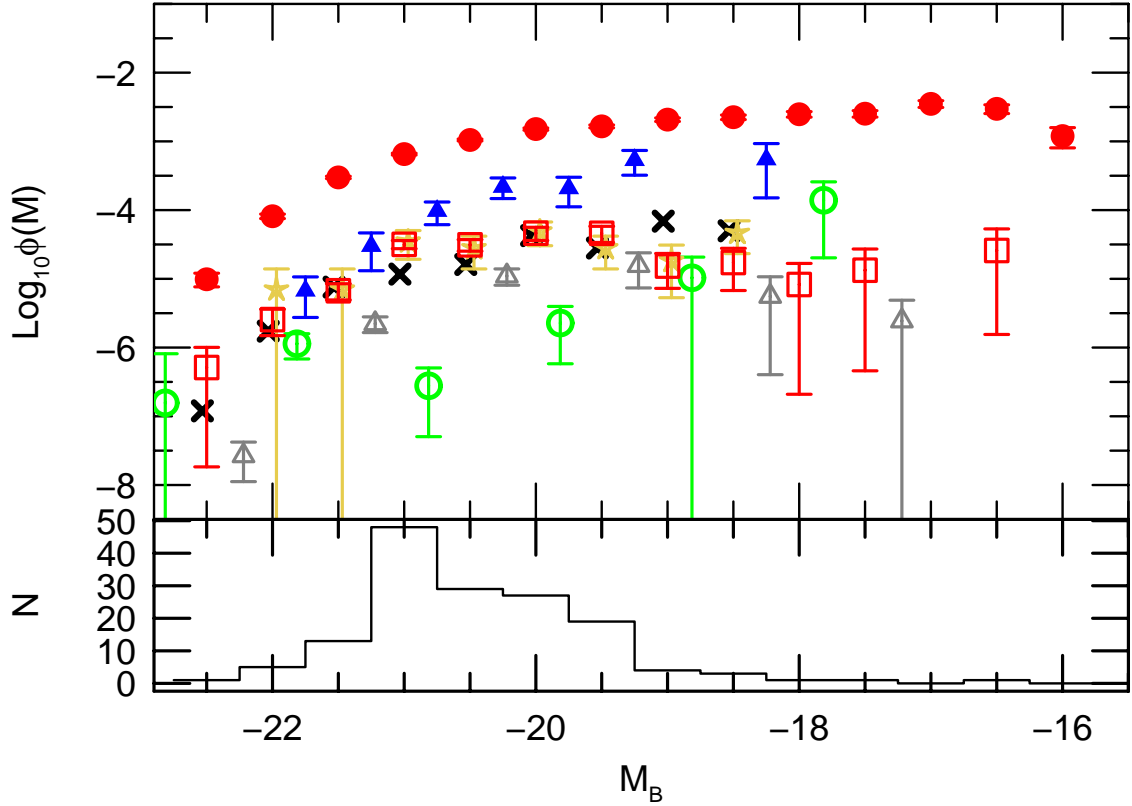}
\vspace{3cm}
\caption {AGN Luminosity function for several low-redshift samples 
of galaxies. The non-parametric stepwise maximum likelihood luminosity function is
shown as filled squares for the AGNC and filled circles for the SSRS2 parent sample. 
Also shown are previous determinations of the luminosity function: \citet{Huc92} - 
crosses; \citet{Koh97} - open circles; \citet{Lon00} - open triangles; \citet{Ulv01} 
- solid triangles; and \citet{Gro01} - stars.}
\label {f11} 
\end{figure}

\clearpage

\begin{figure}
\includegraphics{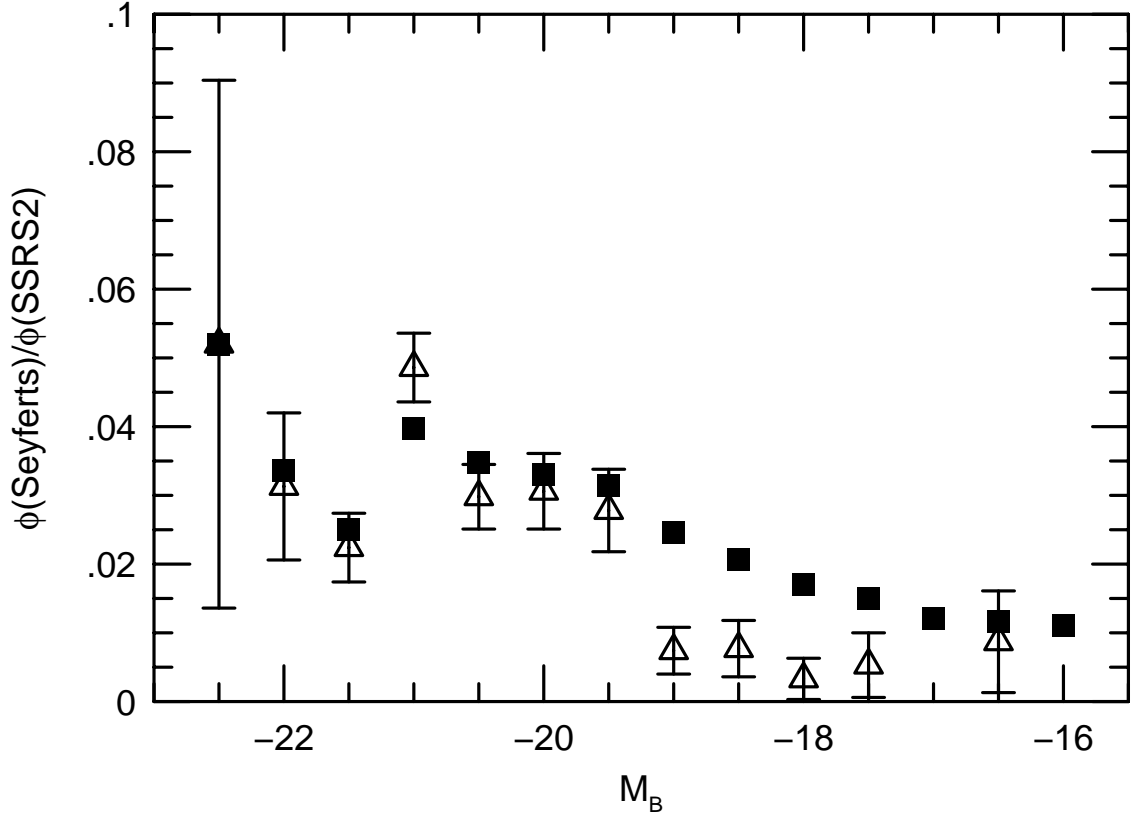}
\vspace{3cm}
\caption {Ratio between the number densities per 0.5 magnitude bin 
of AGN hosts and the parent sample. The open triangles (with associated error bars) 
show the ratio considering only the galaxies within a given bin. The solid circles 
show the cumulative ratio where galaxies in a given magnitude bin and brighter are 
added. The figure suggests there is a trend of AGN hosts being commoner in higher 
luminosity galaxies.}
\label {f12} 
\end{figure}

\clearpage

\begin{figure}
\includegraphics{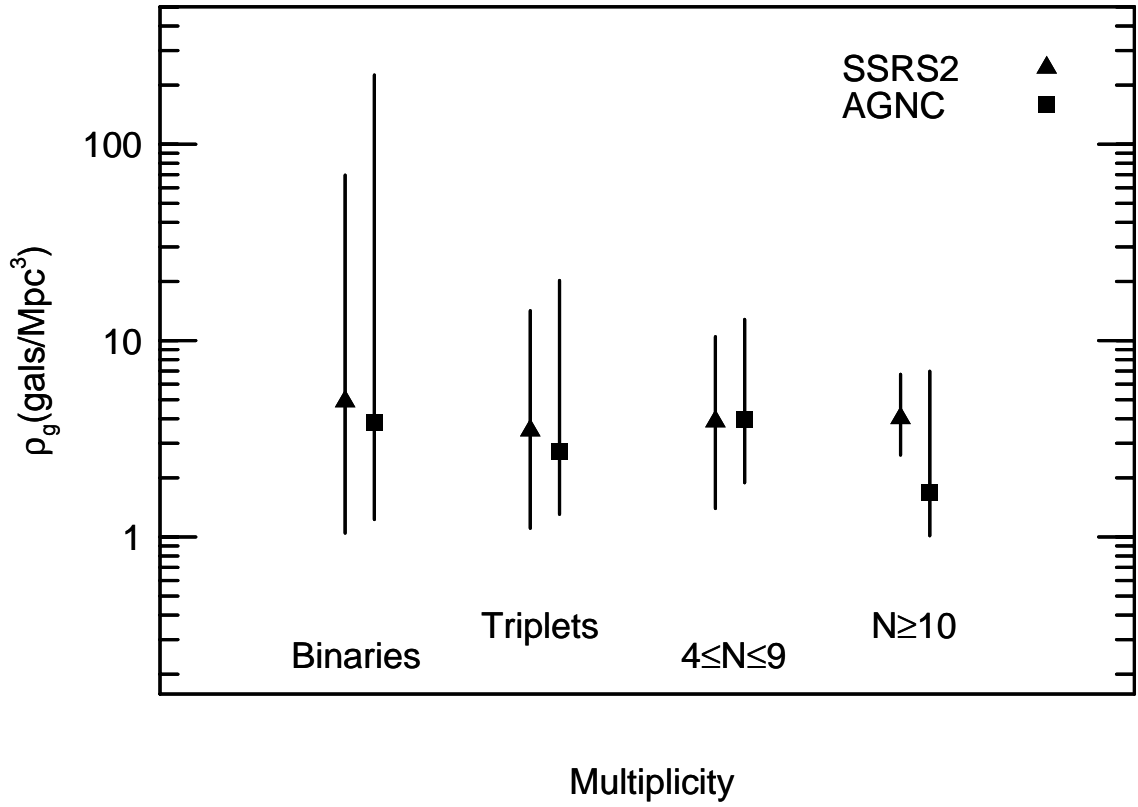}
\vspace{3cm}
\caption {Group density, $\rho_g$, according to multiplicity 
intervals. The points represent median values, while the bars refer to the upper and 
lower quartiles of the distributions.}
\label {f13} 
\end{figure}


\end{document}